# Discerning Authorship in Online Health Communities: Experience, Trust, and Transparency Implications for Moderating AI

YEFIM SHULMAN*, Erasmus University Rotterdam, Netherlands

AGNIESZKA KITKOWSKA, Jönköping University, Sweden

MARK WARNER, University College London, UK

For online health communities, community trust is paramount. Yet, advances in Large Language Models (LLMs) generating advice may erode this trust, especially if users cannot identify whether LLMs have been used. We investigate the feasibility of community-based detection of health advice authorship and how self-moderation of LLMs could help enhance advice utilization. In an online experiment, we evaluate people's ability to distinguish AI-generated from human-written advice across two health conditions, considering lived experience with a condition, AI-recognition training, and user attitudes towards transparency and trust around AI use. Our results indicate the need for transparency coupled with trust. We find little evidence of people's ability to discern advice authorship. However, we find a consistent effect of the health condition. Our qualitative findings identify unreliable signals, resulting in flawed heuristic evaluations of the advice. Our findings point to opportunities to improve the self-moderation of LLM-based AI and aid community-based AI moderation.

## 1 Introduction

People are increasingly using online health communities to complement support received from offline healthcare providers, using these online spaces to seek information and advice and to receive social and emotional support [6, 7, 9, 12, 29, 40, 62, 66]. Online health communities differ from other types of communities as the knowledge shared by those with lived experience of a condition in such contexts can have a critical impact on the lives of others [40], and in helping to shape health-related identity [63]. At the same time, we are seeing a growing interest in the use of generative AI (or Large Language Models, LLMs) across different sectors, including healthcare, in supporting direct-to-consumer health advice [3, 17, 34, 36]. Yet, there are concerns related to risks from these technologies, especially across online social spaces where we are already seeing generative AI being misused, from deepfake text, images and videos [22, 65] to the production of misinformation [88]. Within the healthcare domain, there are concerns related to the quality of advice [3, 74], the risk of health misinformation [17, 74], and the potential risk of harm that these systems create [43].

Where health advice is being provided, it is important to users that the authorship of advice is made available to help maintain trust within online health communities [32], as prior work shows how trust is affected by the use of AI within certain contexts [38]. The transparency around the use of AI can help maintain trust, as the two constructs are often related [42, 48, 56, 68, 78]. Drawing on a dictionary definition [8], we understand *transparency* as the quality of a system or situation such that the underlying processes and activities are performed in an open way. Online content moderation mechanisms, and self-moderation by LLM-based AI could be used to enhance transparency around online

---

*Corresponding author

Authors' Contact Information: Yefim Shulman, shulman@essb.eur.nl, Erasmus University Rotterdam, Rotterdam, Netherlands; Agnieszka Kitkowska, agnieszka.kitkowska@ju.se, Jönköping University, Jönköping, Sweden; Mark Warner, mark.warner@ucl.ac.uk, University College London, London, UK.

health advice, consequently providing the advice's provenance and helping users to evaluate its credibility; however, developing reliable mechanisms for moderating such content is challenging.

In the area of online content moderation, there has been an increasing interest in the use of community-based approaches, such as X's (formerly Twitter) community notes feature that allows the community to write and vote on notes that aim to add context to posts. An advantage of this approach is in the way it brings the community into the moderation process and draws on its expertise; this is an approach that could be used in helping to evaluate the authorship of online health information by those with lived experience of the health condition to which the information relates. Yet, little is known about how lived experience and basic training to detect AI-generated text might affect people's ability to discern authorship, despite prior research suggesting that the inclusion of "experts" into the authorship evaluation process may help to improve detection rates [76].

Another moderation opportunity is self-moderation by LLM-based AI. Most LLMs implement self-moderation mechanisms to prevent certain content from being generated, such as socio-politically biased information [24]. Self-moderation may also detect where certain information is requested, and embed text-based warnings to safeguard users and enhance transparency, for instance, where health advice is sought (Figure 1). An issue with this approach is the ease with which these embedded safeguards can be removed. People viewing information would need to rely on the generated text without these warnings to help them discern authorship. Yet, prior research shows people typically perform quite poorly at detecting AI-generated texts across different contexts and information types [10, 39, 44, 46, 58] which may be due to people using flawed heuristics (e.g., grammar use) embedded within the text [39] that are amplified within computer-mediated communication environments due to a lack of non-verbal cues [79]. LLMs that generate human-like health advice may include unreliable signals that people ordinarily use within heuristic evaluations to discern AI-generated advice.

I'm not a doctor, but I can try to provide some general information. If you have a sore head and a nosebleed, there could be several potential causes. Here are a few possibilities:

Fig. 1. Example of an embedded text-based safeguard ("I am not a doctor...") advising users that the model is not a medical professional.

To understand the feasibility of community-based detection of AI-generated health advice and how self-moderation of LLMs could help enhance the utilisation of such advice, this research explores the interplay of transparency and trust in the process of discerning AI-generated text within a healthcare context. We investigate how effective people are at discerning AI-generated health advice across Health conditions, considering the effects that Training and Lived experience with a given Health condition might have on users' ability to discern between AI-generated and human-written text. This could help support community-based approaches to content moderation and understand the characteristics within the text that people use to support their evaluation of the advice, which could help to improve the self-moderation of LLMs through the removal (or addition) of certain linguistic cues to help people detect AI-generated advice.

Findings from our mixed-methods study show that people want transparency around the use of AI-generated texts within online health community contexts, and this correlates highly with trust levels. When evaluating how effective people are at distinguishing between AI and human-written text, we found no effect of Lived experience or Training. However, we found a difference between the two Health conditions (Diabetes and Back pain). Supporting prior work [38], we found people using heuristics within their decision-making, and, in the Diabetes condition, these heuristics tended to be more flawed.

Our work makes the following contributions:

- Our findings around transparency, considering its division into societal and private organisations delivering the advice, and how transparency relates to trust, support the growing interest in research on the LLMs' provenance. In particular, we contribute by identifying the information around the advice provenance that should be presented in a transparent manner in the context of online health advice.
- We contribute to the growing literature on AI authorship detection within the context of online health communities. In doing so, we offer insights into the limitations of a community moderation approach in detecting AI-generated advice, even where training and lived experience are present.
- We add further support to Walther's [79] hyperpersonal model of communication and Donath's [13] signalling theory, using these theories to interpret our findings and present novel implications for self-moderation of LLM-based AI to support community-based detection of AI-generated health advice to enhance transparency and trust.

## 2 Related work

### 2.1 Information credibility within online health communities

Where information and advice are sought from online health communities [29, 47, 60, 62], the credibility of information shared can impact levels of trust within these communities. While prior studies show how reputable sources, such as government and hospital websites, can help enhance trust and reliability in information, both O'Kane et al. [60] and Burgess et al. [6] found people with complex chronic conditions often lacked trust in information received from their clinicians due to a lack of trust in the level of expertise that their clinicians had about their complex chronic conditions. Their research identified people using online health communities to seek out patient-generated sources of information to help them make sense of their condition, a behaviour we also see around Polycystic Ovary Syndrome (PCOS) [29], pregnancy [47], male factor infertility [62], Chronic Kidney Disease (CKD) [6], and HIV [7]. Yet, people often struggle to evaluate the credibility and reliability of online information and advice, or lack trust in such information [47]. In Gomula et al. [29] study into how women with PCOS seek information from online communities to make sense of their condition, they found women reporting a lack of trust in some information sources, especially those that were looking to commercialise PCOS advice; a finding supported in other health contexts, such as pregnancy [47]. While some prior work finds advice from online health communities to be more trustworthy and highly valued [62], where online health advice is considered less reliable, people use different approaches to evaluate its veracity. For example, using search engines or other health information sources [29]. Yet, Burgess et al. [6] highlight the challenges people face in evaluating health information online, suggesting the need for better resources and education to support people in these processes.

### 2.2 Trust and transparency around AI-generated health information

Trust can be defined as "willingness of a party to be vulnerable to the actions of another party based on the expectation that the other will perform a particular action important to the trustor, irrespective of the ability to monitor or control that other party" [55, p. 712]. Trust is a factor important for human-to-human relationships, but also for relationships with non-humans, such as with AI-based technologies [27] that can be used to generate content within online communities. Expert evaluations of AI-generated health advice have demonstrated a complex interplay between trust and transparency. In a study by Van Bulck and Moons [77], experts assessed the trustworthiness and safety of health advice generated by ChatGPT. The health advice was deemed trustworthy and potentially valuable for patients due to the comprehensibility of the responses and the tendency of the AI to refer patients to healthcare specialists. However, some experts criticised the advice in terms of its writing style (e.g., not patient-centric), and for being of low quality (e.g., missing information) [77]. Despite the increasing adoption of AI-based technologies, several studies indicate that people might not trust these types of systems. A recent survey indicated that less than 40% of approximately 17,000 respondents from 17 countries expressed trust and acceptance of AI-based systems. However, higher trust was observed in healthcare applications (44%) and amongst participants from certain non-Western geographical regions [28].

Transparency (as the quality of a system or situation such that the underlying processes and activities are performed in an open way) around the use of AI is critical within health systems and can help maintain trust, as the two constructs often relate [42, 48, 56, 68, 78]. Much of the emerging work on transparency of LLMs focuses on their "blackbox" nature, as information on how models are trained, and the type of data they are trained on are rarely made public [?]. Moreover, many of these systems fail to communicate justification to support their generated output [72, 81], a significant concern when the output relates to health advice and diagnostic information [1?]. While these aspects of transparency are important, within the context of ethics of AI, there is also a recognised need for companies developing and designing AI-based technologies to "Be transparent about the fact that they are dealing with an AI system" [35]. Perceived transparency is further related to a number of factors — for instance, it may be positively correlated with the perceived quality and interpretability of an AI-powered recommender platform [70]. Despite the importance of these interrelated user perceptions and attitudes, research on the role of trust and transparency in AI-generated health advice remains scarce. Given the important role that online health communities have across a spectrum of health conditions, the increasing use of LLM-based AI technologies within healthcare, and the lack of research into people's attitudes around the use of generative AI within the context of online health communities, we pose the first research question (as part of our conceptual research model shown in Fig. 2):

**RQ1:** What are the relationships between people's attitudes towards transparency and trust in AI-generated advice within online health communities and the ability to discern between human-written and AI-generated health advice?

## 2.3 Reliability of signals and discerning AI-generated content in online health communities

While trust is a critical component when interacting online, especially within the context of health, it can be challenging to develop trust in online communities, as these computer-mediated communication (CMC) environments typically lack the contextual and social cues that are present in face-to-face communications. Walther's [79] hyperpersonal model of CMC argues that in the absence of social cues, more subtle cues (e.g., spelling errors) are amplified and used by receivers to help them develop perceptions of others; behaviour that has been observed in empirical studies across different contexts including online dating platforms [18], social networking sites [75], and online health communities [20, 52]. Donath [14] applied aspects of signalling theory from evolutionary biology to the area of CMC, suggesting that people develop and use signals (intentional communications of quality) and cues (inferences of a hidden quality) to develop more reliable perceptions of others online. They describe signals that are difficult to manipulate (e.g., verified account status on X (formerly Twitter)) or costly to produce (e.g., the disclosure of a socially costly stigmatising condition [82, 83]) as 'honest signals'.

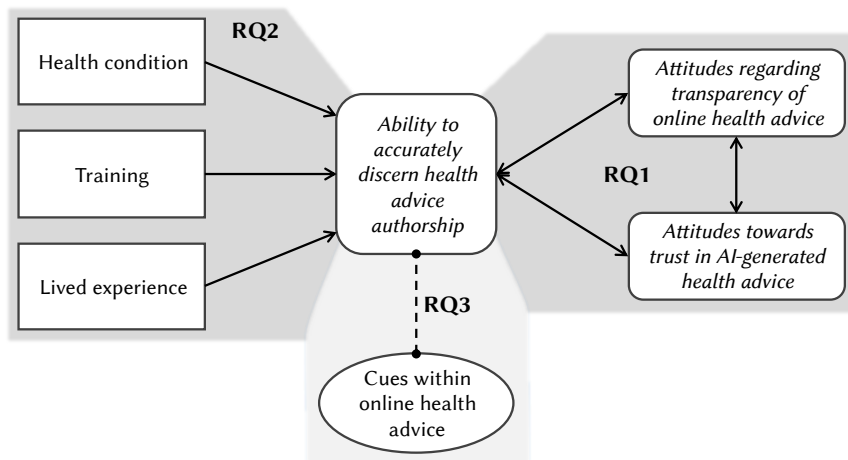

Fig. 2. Research model

Aligned with these theories, prior work [20] suggests people develop trust in online health communities through the evaluation of the credibility of information within individual posts and the longer-term evaluation of both the content of posts, and the sources of those posts (i.e., the posters). When evaluating information credibility, Fan et al. [20] identified heuristic cues (e.g., grammar, quality of sources, the competence of the poster) and cognitive assessments (e.g., the logic of the advice) being used by people to help them infer the trustworthiness of the poster and their advice. Other work [52, 52, 53, 62, 64] has explored the practice of collective sense-making within communities, to help collectively evaluate, understand and accept information and advice shared within the community, often using various cues within or associated with posts to help in this process. However, the advent of generative AIs such as LLMs risks limiting the amount of information that can be used to *warrant* [11, 80] through its ability to produce content that can appear human-created, resulting in less reliable (or 'dishonest' [13, 14]) signals. Recent empirical work has explored the relationship between trust formation and the use of AI to generate self-presentations (i.e., user profiles) within the context of an online marketplace [38]. This work found mistrust developing around profiles generated, or suspected to have been generated, using AI; this effect was only present where AI profiles existed in environments alongside human-written profiles, developing negative impressions of the profile owner based on minimal cues within the profile text. Drawing on this and other work, Hancock et al. [32] define a new form of communication mediated through AI agents. Hancock et al. [32] posit that AI-mediated communication (AI-MC) requires us to reevaluate many of our existing theories of CMC such as the hyperpersonal model [79] and signalling theory [14], due to the ease with which generative models can manipulate signals and cues within these environments. Our work responds to the research agenda proposed by Hancock et al. [32] by exploring factors that might result in people being less susceptible to manipulation by LLMs and, in doing so, identify factors that may result in increased manipulation susceptibility. Where the presence of AI-generated content has the capacity to reduce levels of trust within communities, understanding users' attitudes towards transparency of use of LLM-based generative AI, and their ability to discern between AI-generated and human-written content is critical due to how these elements interrelate with trust.

2.3.1 *Discerning between human and AI-generated content.* Due to AI's potential to impact trust in online communities [32, 38], social computing researchers are increasingly interested in studying and understanding how, and how well people discern between human and AI-generated content such as text [10, 39, 44, 46, 58, 61], images [5, 31, 61], audio [51, 57], and video [30, 33, 45, 73]. Although evaluating the ability to recognise machine-generated output dates back to Alan Turing's imitation game, recent advancements in generative AI have heightened this focus. This is not only due to the increased accessibility of text-based generative AI systems but also due to the enhanced human-like characteristics of contemporary LLM-based AI models, which can assume different roles and provide contextually nuanced responses (e.g., ChatGPT acting as a helpful assistant) [69].

Advancements in LLMs are making it harder for people to accurately discern between human-written and AI-generated content. In an online experimental game, in which approximately 2k participants were asked to discern between human-written and AI-generated content (text and images), Partadiredja et al. [61] found that, on average, about half of the responses were incorrect. Only 45% of images were attributed correctly, while for texts, the score was slightly higher—54%. Clark et al. [10] investigated the recognition of human vs. machine-generated texts and the Training effect in three different text-based contexts (newspaper, stories, and recipes). Their results indicate that people's ability to discern is not increased significantly even after training evaluators in recognising machine-generated text. Moreover, their study indicated that people tend to underestimate the abilities of AI and believe that machines are unable to create responses that possess human characteristics. Jakesch et al. [39] found that people's inability to accurately discern AI-generated self-presentation authorship was partly due to flawed heuristics (e.g., grammar use) being used by evaluators. Although it is common to assume that heuristic-based decisions can be flawed, some argue that this is not always the case. For instance, Gigerenzer and Gaissmaier [26], in their overview of non-rational theories of decision-making, presented a classification of heuristics, showing that when they are used in specific environments or by specific individuals, professions, or groups, heuristic-based decisions outperform those based on complex statistical predictive models. Yet, heuristics often rely on information that is reliable or 'honest' [14]. Where signals are dishonest, they may be used in heuristic-based decision-making, resulting in heuristics being flawed, as seen in prior work [32, 39].

Within the context of health, there has been little prior work exploring the discernment of AI-generated information. However, one study has explored the distinction between chatbot-generated and human-generated health responses. Nov et al. [58] found the accuracy rates for distinguishing chatbot responses and human responses to health queries to be very close, at 65.5% and 65.1%, respectively. These results underscore the subtlety of differences and the potential for generative AI to convincingly emulate human communication, which can result in flawed decision-making heuristics. To counter this, prior work has evaluated the effect of training users on how to spot AI-generated content. Training has been found to be effective in helping users detect AI-generated video [73], yet around text-based content, prior work shows its effect is dependent on information topic [10]. As such, we evaluate the effect of Training within an online health information context.

There has also been very little prior work to understand how domain knowledge of the content being evaluated impacts the ability to discern AI-generated from human-written text. As people within online health communities often have lived experience of the condition that the community is centred around, we explore the effect of Lived experience (i.e., domain knowledge) on discerning between AI-generated and human-written advice. Additionally, as health conditions attract diverse semantics and differences in the complexity of health terms used within online health communities [85], we also explore how the medical condition itself impacts discernment efficacy. To investigate these variables, we pose the following question:

**RQ2:** How do Training, Lived experience, and the type of Health condition impact the ability to accurately discern between human-written and AI-generated health advice?

To investigate the RQ1 and RQ2, being informed by prior literature, we develop a set of research hypotheses (Table 1) for empirical testing (Section 3).

Table 1. Summary of research hypotheses (addressing RQ1 and RQ2).

| Label | Hypothesis | Rationale |
| --- | --- | --- |
| RQ1.H1 | Transparency attitudes towards health advice positively correlate with trust in AI-generated health advice. | Conceptual and empirical relatedness of the constructs |
| RQ1.H2 | Transparency attitudes correlate with the ability to discern health advice authorship. | Potential relatedness to perceived advice quality |
| RQ1.H3 | Trust in AI-generated health advice correlates with the ability to discern health advice authorship. | Potential relatedness to perceived advice quality |
| RQ2.H1 | People who received Training on how to identify AI-generated health advice are detectably better at discerning the advice authorship, than those who did not receive such a Training. | Altering decision-making approach and heuristics |
| RQ2.H2 | People who have Lived experience with the health condition, to which the health advice is addressed, are detectably better at discerning the advice authorship, than those who do not have such a Lived experience. | Utilising first-hand, relevant domain knowledge |
| RQ2.H3 | There exists a detectable difference in the ability to discern health advice authorship between different Health conditions. | Testing complex difference in semantics, contexts |
| RQ2.H4 | People who both received Training and have Lived Experience are detectably better at discerning the advice authorship, than those who have either Training or Lived Experience. Those who have either Training or Lived Experience are also detectably better than those who have neither Training nor Lived Experience. | Altering decision-making and utilising domain knowledge may reinforce reciprocally |

Finally, although Jakesch et al. [39] found heuristics for discerning AI-generated language are flawed, this has not been explored within the context of health advice. This is particularly pertinent as prior work has shown how people use cues and assessment criteria to evaluate trust in information and its source within online health communities [20]. As such, we pose the following research question (Fig. 2):

**RQ3:** What cues do people use within health advice to discern authorship?

## 3 Method

To better understand the feasibility of community-based detection of the provenance of online health advice and how self-moderation of LLMs could help enhance the utilisation of such advice, we posited a research model (Fig. 2) illustrating and relating the research questions. To follow the research model and address the research questions, we developed a mixed-methods online study, recruiting participants (Table 3) through the online recruitment portal Prolific. We used pre-screening based on the Lived experience with our two chosen Health conditions — Diabetes and Back pain (we elaborate on this choice in Section 3.3). The study was a controlled online experiment with a 2 × 2 × 2 between-subjects experiment design (Table 2), including questionnaires that measured people's attitudes towards transparency and how they relate to trust in AI-generated advice within online health communities (RQ1). In the main task, we showed participants health advice across two Health conditions and asked them whether the advice was AI-generated or human-written (RQ2). Finally, we embedded open-ended questions to capture and analyse the reasons people gave for their decisions on advice authorship (RQ3).

Table 2. Independent variables in the experiment (addressing RQ2).

| Variable | Level | Participants, $n$ |
|---|---|---|
| 1. Training | A. Present (alongside familiarisation with the task) | 127 |
|  | B. Absent (familiarisation only) | 125 |
| 2. Lived experience | A. Present | 127 |
|  | B. Absent | 125 |
| 3. Health condition | A. Diabetes | 125 |
|  | B. Back pain | 127 |

*Note:* In the study, we had from 29 to 34 participants per independent group, $M = 31.5$, $SD = 1.51$, for the full sample of $N = 252$.

### 3.1 Participants

We used Prolific to recruit participants, paying them an average of GBP19.74/ph (screening) and GBP12.88/ph (main study) [1]. Overall, we recruited N=612 UK participants in the pre-screening stage. We invited N=505 participants to the main study, from which we obtained N=257 complete responses (Table 3). We removed respondents who completed the study in less than five minutes, retaining N=252 valid responses for the analysis. The sample size was sufficient as per a priori power analysis performed in G*Power 3 [21]. The participants' demographics included in the analysis are shown in Table 3. Most participants did not experience any cognitive difficulties (96.4%, $n = 243$). Seven participants reported having some difficulties: two of them having learning difficulties, four — reading difficulties, and one participant having both learning and reading difficulties. Two participants preferred not to say. In terms of ethnicity, 92.9% ($n = 234$) of participants self-identified as "White", $n = 7$ indicated "Black or Black British", $n = 6$ selected "Asian or Asian British", and $n = 5$ indicated "Mixed or multiple groups".

### 3.2 Materials

*3.2.1 Questions, advice, and model prompts.* We sourced the questions and human-written answers from Reddit's r/AskDocs community discussions as *"All flaired [tagging, trusted] medical professionals on this subreddit are verified by the mods"*. Having considered ethical implications, we included these materials in the study only after we removed personal details that Reddit users may have shared in their questions and answers. We ensured that our modifications did not change the substantive and significant elements of the questions. We used these modified versions of the questions as prompts to obtain AI-generated replies through ChatGPT to avoid making personal data available to the algorithm for storage and processing.

Prior to generating responses, we refined the prompt parameters within the *GPT-3.5-turbo* model. First, we set the system role from the default *"You are a helpful assistant"* to *"You are a medical doctor and are replying to a health-related question that a user has posted on an online forum"*. Moreover, we instructed the response length to be limited to improve the consistency of advice length between the AI-generated and human-written advice. The *temperature* parameter controls the probability distributions for the tokens of the model response. Higher *temperature* values result in a more "creative" yet varied response that can reduce the coherence of the model's output. Lower values lead to less "creative" and more deterministic responses that may usually be more coherent. We set the *temperature* parameter to a midpoint value of 0.5 for a balanced determinism of the model. The *top_p* parameter (nucleus sampling) controls how many possible words to consider within the generated response. A high value causes the model to consider more possible words, making the generated response linguistically more diverse. We set the *top_p* parameter to 1.0, which, combined with a midpoint *temperature*, provides a more coherent, yet "creative" response with a more diverse linguistic output. Both the *frequency_penalty* and *presence_penalty* control the frequency of word use within the model response. As we were only generating a single response, word frequency was less of a consideration, so both of these parameters were set to a mid-range point of 0. Finally, we removed any text-based safeguards that were prefixed or suffixed to the generated advice, such as *"I am not a doctor, but I can try to provide some general information"* (Figure 1).

---

[1]Payment for the screening survey was higher as we overestimated how long the study would take, whereas, for the main study, our estimation was more accurate as we were informed through pilot studies.

Table 3. Participants' demographics, $N = 252$.

| Demographic | Level | n | % |
|---|---|---|---|
| Gender | Female | 126 | 50 |
| | Male | 125 | 49.6 |
| | Non-binary or self-described | 0 | 0 |
| | Preferred not to say | 1 | 0.4 |
| Age cohort | 18–24 | 6 | 2.4 |
| | 25–34 | 65 | 25.8 |
| | 35–44 | 71 | 28.2 |
| | 45–54 | 44 | 17.5 |
| | 55–64 | 46 | 18.3 |
| | 65–74 | 16 | 6.3 |
| | 75–84 | 4 | 1.6 |
| | 85 or older | 0 | 0 |
| | Preferred not to say | 0 | 0 |
| Highest completed level of education | Less than high school | 1 | 0.4 |
| | High school or equivalent | 49 | 19.4 |
| | Some college or university | 63 | 25 |
| | Bachelor's degree or equivalent | 90 | 35.7 |
| | Master's degree or equivalent | 38 | 15.1 |
| | Doctorate degree or equivalent (e.g., DPhil, PhD) | 9 | 3.6 |
| | Other, please specify | 2 | 0.8 |
| | Preferred not to say | 0 | 0 |
| Employment status | Unemployed | 13 | 5.2 |
| | Employed (full-time, part-time, or casual) | 183 | 72.6 |
| | Self-employed | 12 | 4.8 |
| | Full time student | 5 | 2 |
| | Retired | 26 | 10.3 |
| | Other | 12 | 4.8 |
| | Preferred not to say | 1 | 0.4 |
| Total | | | 100.0 |

For the controlled experiment (Figure 4), we selected three questions for each Health condition and prepared one human-written and one AI-generated reply per question. For the familiarisation/Training sub-stage, we selected three health questions unrelated to Diabetes and Back pain, and prepared one human-written reply for one question and one AI-generated reply for each of the other two questions. All questions and replies can be found in the supplemental materials.

3.2.2 *Training to identify AI-generated text.* To train participants in how to identify AI-generated advice, we adapted approaches from prior research [10, 15]. Clark et al. [10] evaluated three approaches to training, including training with instructions, with examples, and with comparisons (e.g., compare human with AI). They found training with examples to be the only method that showed a significant improvement in people's ability to discern, and so we utilised the training with examples approach within the present research. Further work has been conducted to help improve training with examples. Dou et al. [15] developed a framework consisting of categories of commonly occurring issues found in AI-generated text; these categories are used to annotate text [15]. While we did not annotate the text of our main study examples, we applied this framework and this annotation method to our training examples (Figure 3).

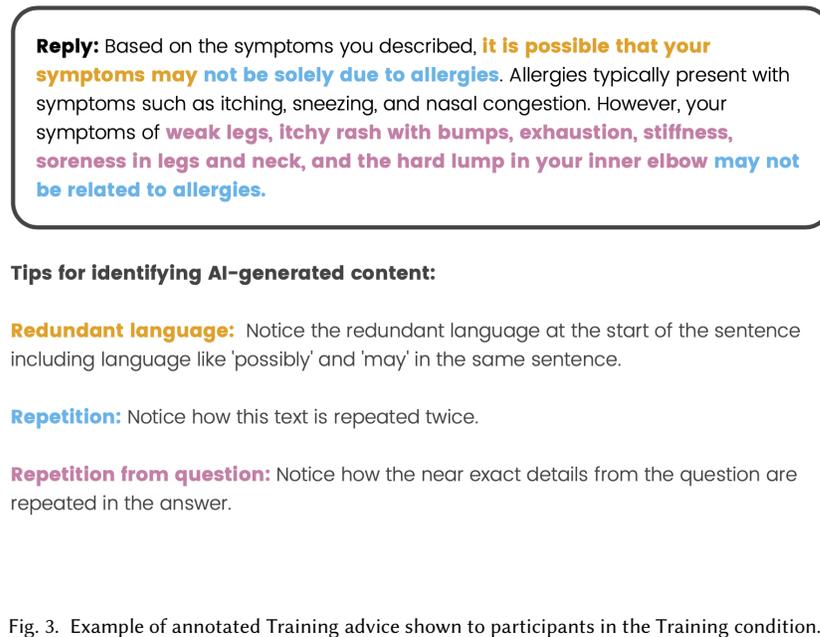

Fig. 3. Example of annotated Training advice shown to participants in the Training condition.

*3.2.3 Transparency and trust measures.* To measure participants' attitudes towards transparency around health advice on online social platforms, we used a set of ad-hoc statements (Appendix A.1.1) created following our definition of transparency (Section 1). To measure participants' trust towards AI-generated health advice, we used a scale adapted from Nov et al. [58], where we modified the statements to be relevant to our topic (Appendix A.1.2). The transparency and trust sets contained six items each, with one of the transparency items containing six statements. Finally, we included an additional ad-hoc item regarding participants' comfort of receiving AI-generated health advice. All statements used 7-point Likert-type items, anchored "strongly disagree" through "strongly agree". Exact statements can be found in the supplementary materials.

## 3.3 Study Design

*3.3.1 Health condition.* The selection of two Health conditions — Diabetes and Back pain — had a three-fold motivation. Firstly, we wanted to ensure that the health conditions were common enough that we could reach the desired sample size. It is estimated that in the UK, there are 5.6 million people living with diabetes [2], and approximately 11 million people in the UK experience back pain, of which 6.5 million experiences severe lower back pain [2]. Second, the screening of Reddit's r/AskDocs resulted in these two topics being discussed frequently on the platform, enabling a larger poll for a selection of questions and advice. Lastly, the two conditions might be perceived as opposing — diabetes is commonly perceived as a long-term condition that requires permanent treatment, while back pain might be perceived as a temporary condition that does not require regular treatment and is of a more "trivial" nature.

---
[2]https://www.diabetes.org.uk/about-us/about-the-charity/our-strategy/statistics

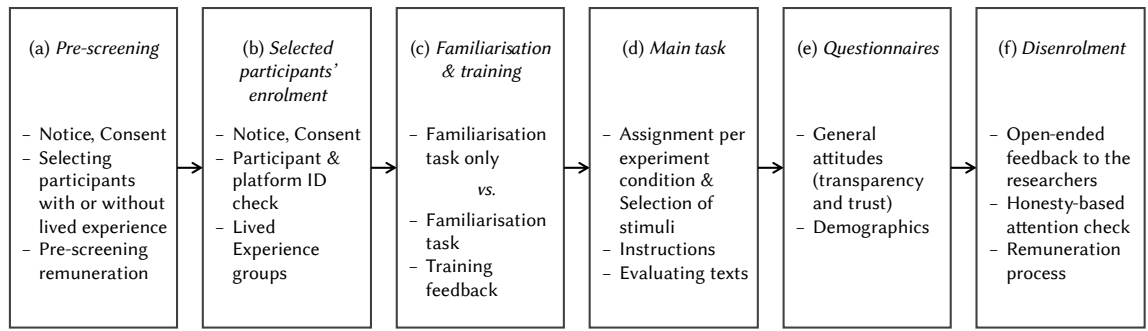

Fig. 4. The study flow.

*3.3.2 Training and familiarisation.* Participants in all conditions were required to perform the assessment task across the Training advice, however only participants in the Training group were informed whether they had assessed authorship correctly, and were always shown the annotated version of the advice (Figure 3). Those in the Non-Training group were required to complete the assessment task across the Training advice as a familiarisation exercise, and were not informed whether they had assessed authorship correctly and were never shown the annotated version of the advice.

*3.3.3 Outcome measures.* We collected quantitative and qualitative outcomes from our study design. The measures that we applied in the study include:

(1) Participants' evaluation of whether the reply (i.e., advice, post, answer) was human-written or AI-generated on a 4-point scale [10, 37, 38]. Participants indicated whether the reply, in their opinion, was "definitely human-written", "possibly human-written", "possibly AI-generated", or "definitely AI-generated" (Figure 5). This measure is used in the quantitative analysis (Section 4) to obtain the dependent variables reflecting *correctness* of participants' evaluations and to address RQ2.
(2) Participants' attitudes towards transparency in the context of receiving online health advice on social platforms and their trust in AI-generated health advice (RQ1), collected through Likert-type items (Section 3.2 and Appendix A).
(3) Participants' explanations of their reasoning behind each of their evaluations, collected through open-ended questions (Figure 5). The answers provided data for the qualitative analysis addressing RQ3.

*3.3.4 Procedure.* The stages of the study are shown in Figure 4 and include the following:

(a) *Pre-screening.* We asked participants whether they had lived experience with either of the two Health conditions or lived with someone (e.g., a partner) with either of the conditions.
(b) *Selected participants' enrolment.* Upon reaching a sufficient number of participants per each between-subject group, we invited qualifying participants into the main study. We considered people to have Lived experience if they were living with the condition, or were living with someone (e.g., partner, relative) with the Health condition, as living with someone with the condition may also be considered a form of Lived experience. We considered people to have no Lived experience if they were not living with the condition, nor living with someone (e.g., partner, relative) with the Health condition.

(c) *Familiarisation only vs. familiarisation and Training.* Participants assigned to each group combination of Lived experience and Health condition were randomly divided into two equal groups: one went to familiarise themselves with the experiment task (Training—Absent); whereas the other group were trained on how to recognise AI-generated text (Training—Present). Both groups were exposed to the same questions and replies, but in addition to familiarisation with the evaluation process from the main task, the trained group received feedback on how they performed and hints on what to look for in trying to identify AI-generated texts, based on prior approaches [10, 15]. Participants in both groups provided their evaluation in a manner identical to the one used in the main task — Figure 5 shows the evaluation interface. Participants were not able to skip either the familiarisation or Training task.

(d) *Main task for all participants.* All participants were shown three pairs of questions and replies. Each question was followed either by a human-written reply (from a health physician on Reddit) or a reply generated by LLM-based AI (ChatGPT 3.5). The type of reply was randomised for each participant so that the replies were shown an approximately equal number of times for each question. No question was repeated within-subject, so participants would not be able to compare the two replies to the same question when providing their evaluations (i.e., was the response generated by human or AI). As a part of the evaluation (Figure 5), participants were asked open-ended questions after they submitted their quantitative evaluations. In line with prior research [10], participants were asked, *"Why do you think it is definitely human-written?"* if they indicated it was definitely human-written, otherwise, they were asked, *"What would you change to make it seem more human-like?"*.

(e) *Questionnaires.* Participants answered questions measuring their attitudes towards transparency when dealing with online health advice and their trust in and comfort with AI-generated health advice. The order of statements was randomised separately for the two sets of attitudes. We also collected demographic information (age cohort, gender, education, and learning and reading disabilities).

(f) *Disenrolment.* Finally, participants were invited to share with the researchers any feedback regarding the study and their participation. They were also shown a debriefing message corresponding to their experimental group and were redirected back to the experimental platform to receive their remuneration.

*3.3.5 Study ethics.* Ethical considerations were embedded into our research discussions and impacted our decision-making around the design of our study. We considered the impact our study might have on our participants, and aspects related to data protection and privacy around data we were collecting from participants, from online sources, and data we were inputting into ChatGPT. Our study was approved under the [anonymised] departmental ethics approval procedure (ref: anonymised).

### 3.4 Data analysis

*3.4.1 Quantitative analysis.* The quantitative data obtained through the experiment were analysed with different statistical tests. Descriptive analyses (correlation analysis, exploratory factor analysis) were applied to better understand the data and answer RQ1. Analysis of covariance (ANCOVA) models were used to answer RQ2. Before the data analysis, the test assumptions (normality, homogeneity) were checked. Although slight violations were identified, the parametric

## (a) Advice assessed as human-written.

## (b) Advice assessed as AI-generated.

Fig. 5. The task interface shown to participants to illicit their assessment of the advice, and their qualitative reasoning.

tests used in the analysis are robust against violations, particularly in larger samples. Therefore, we proceeded with the analysis.

*3.4.2 Qualitative analysis.* In total, we collected 710 qualitative responses to our open-ended questions (Figure 5). The shortest response was 4 characters, and the longest was 170 characters, with a mean length of 57. All authors were involved in an initial round of inductive coding of the data to gain familiarity with the data and to each develop a separate initial codebook. All authors met to discuss their initial codebooks. At this stage, some codes were collapsed. For example, codes "shows empathy" and "shows sympathy" were collapsed into "shows emotion". Once an initial codebook (consisting of 22 codes) had been developed, the first and third authors conducted a further two rounds of independent coding to revise the codebook. In each iteration, they independently coded 10% of the data, after which all authors met to discuss disagreements and revise the codebook to address any coding issues identified. For example, in the first round, we added a code 'Includes irrelevant detail' as this was not captured by our existing codes. Moreover, through these discussions, it was identified that multiple codes were occasionally applicable, so we decided to allow for two codes to be applied to each response. For the second round of independent coding, we calculated inter-rater reliability using Cohen's (unweighted) $\kappa$ to evaluate the reliability of our developed codebook, computing an agreement of $\kappa$=0.71. The first and third authors met again to discuss and resolve disagreements, changing the description of codes to improve code reliability. Having reached a high agreement after discussion and agreed on the suitability of the codebook (consisting of 24 codes), the first and third authors each independently coded the entire dataset (reaching an agreement of $\kappa$=0.73), with all authors reviewing the final codes to resolve remaining disagreements. Finally, the codes were sorted into themes, and the names of the themes were collaboratively revised to describe the codes they represent appropriately.

Table 4. Composition of the obtained transparency variables.

| Transparency variables | Items | Factor loadings |
| --- | --- | --- |
| | I think it's helpful to know if a particular health advice I receive online is reviewed or moderated by: | |
| 1. *Transparency of Societal Moderation* (*Transparency-Societal*): | — Professional association (medical, academic, etc.) | .776 |
| | — National Health Service | .766 |
| | — Non-profit organisation (e.g., health charities) | .477 |
| 2. *Transparency of Private Moderation* (*Transparency-Private*): | — An AI health advice system | .780 |
| | — For-profit organisation (e.g., private health providers) | .661 |
| | — Other members of an online health community | .491 |

## 4 Results

### 4.1 Descriptive statistical analysis of attitudes towards transparency and trust (RQ1)

To better understand participants' attitudes towards transparency when dealing with online health advice, and their trust in and comfort with AI-generated health advice (RQ1), we perform descriptive statistical analyses of the questionnaire measures obtained in the controlled experiment.

As we adapt an existing trust scale [58], we first check whether the items load into a single factor by performing an exploratory factor analysis (EFA). The principal component analysis (PCA) results were satisfactory, with Kaiser-Mayer-Olkin (KMO) of sampling adequacy equal to .88, and Bartlett's test of sphericity significant, $p < .001$. All the trust measurement items loaded into one factor. The reliability of the trust measurement was checked with Cronbach's $\alpha = .90$, indicating high reliability. We used the mean of the items to compute the *Trust* variable.

The series of questions related to transparency was then tested to assess whether it measured one or more latent constructs. Using EFA, with principle axis factoring and varimax rotation, it became clear that not all items loaded into one factor. After the investigation of communalities and removal of all items with communalities lower than two, only six items remained, loading into two factors (three items in each). KMO was .75, and Bartlett's test of sphericity was significant, $p < .001$, indicating the suitability of EFA. Based on one factor, we created a new variable called *Transparency of Societal Moderation* (Table 4), as it included items related to the information about different entities pursuing public or professional interests (professional associations, national health services, or non-profit organisations) as moderators (reviewers) of health advice. We checked the reliability of the construct, and Cronbach's $\alpha$ was good .71. Using the items loading into the second factor, we created a new variable called *Transparency of Private Moderation* (Table 4), as it included items related to the information about entities pursuing individual or private interests (for-profit organisations, AI health advice systems, and members of online communities) moderating health advice. The reliability analysis was good, Cronbach's $\alpha = .70$. To compute the variables (Table 4), we used the mean of items loading into each factor. For short, we may refer to these two variables as *Transparency-Societal* and *Transparency-Private*. The remaining items related to transparency were analysed independently.

We used descriptive statistics to gain insights into the questionnaire data set. In particular, we looked at frequencies of ad-hoc questions and conducted correlation analysis to identify potential significant relationships between the latent constructs and standalone questions. In short, almost half of the participants stated that they would, to some extent, feel comfortable with receiving AI-generated health advice online (47.2%), and 16.3% selected neither agree nor disagree. The majority wanted to know who developed an AI that generates health advice (90.5%). Similarly, 90.5% of

Table 5.  The results of Pearson correlation analysis.

|  | 1 | 2 | 3 | 4 | 5 | 6 | 7 | 8 | 9 |
|---|---|---|---|---|---|---|---|---|---|
| 1. Trust | 1 | | | | | | | | |
| 2. It would be helpful to know what other people think about health advice I receive online | .18** | 1 | | | | | | | |
| 3. It is useful to know if a particular health advice I receive online is financially sponsored by a company or organisation, regardless of whether advice comes from human or AI | .01 | .09 | 1 | | | | | | |
| 4. It is important to me that whoever provides me with health advice online also shares relevant personal and reputational details | −.03 | .16** | .21** | 1 | | | | | |
| 5. If an AI was to provide me with health advice online, it is important to me that it declared information about who it was developed by | −.19** | .14* | .35** | .36** | 1 | | | | |
| 6. I would be comfortable receiving AI-generated health advice online | .81** | .12 | .01 | −.03 | −.19** | 1 | | | |
| 7. Transparency-Societal | −.00 | .25** | .37** | .23** | .32** | −.02 | 1 | | |
| 8. Transparency-Private | .17** | .13* | .13* | −.07 | .08 | .13* | .50** | 1 | |
| 9. Correctness-Overall | .12* | .01 | .01 | .11 | .07 | .12 | .04 | .08 | 1 |

**$p < .01$. and *$p < .05$., $N = 252$.

the participants wanted to know the reputational or relevant personal details of the entity providing health advice. Over half of the participants wanted to know other people's opinions on the online health advice they would receive (60.3%).

The results of the correlation analysis are presented in Table 5. In this section, we describe only significant correlations of medium size or larger. The results indicate a significant strong correlation between Trust and comfort in receiving AI-generated health advice. The positive relationship between these variables suggests that people with higher levels of Trust are more likely to feel comfortable receiving AI-generated health advice online. There is a medium-in-strength correlation between the two transparency constructs (Societal and Private), indicating they may measure related constructs. Moreover, Transparency-Societal is positively correlated with the need for reputational and financial information about sources of online health advice, implying that people preferring transparency about public and professional institutions (societal moderators) might also want to know about their reputation, as well as who sponsors those moderators. Similarly, the measurements around the reputational information and financial sponsorship (columns/rows labelled '3' and '4' in Table 5) positively correlate with the information about who developed an AI health advice system.

### 4.2   Distinguishing between the human-written and AI-generated advice (RQ2)

We first analysed and constructed variables required for the main analysis. We re-coded the participants' evaluations to compute the *Correctness-Overall*. All correct evaluations (e.g., an AI-generated reply identified as "definitely AI-generated" or "possibly AI-generated") scored 1, and all incorrect evaluations (e.g., an AI-generated reply identified as "definitely human-written" or "possibly human-written") scored 0. Then, the outcome variable was created as a sum of correct scores across AI-generated and human-written advice, with Correctness-Overall ranging from 0 to 3. A similar process was used to compute the Correctness-AI and Correctness-Human variables, but considering only responses to stimuli presenting AI-generated and human-written advice, respectively. These two variables scored from 0 to 3, as well.[3]

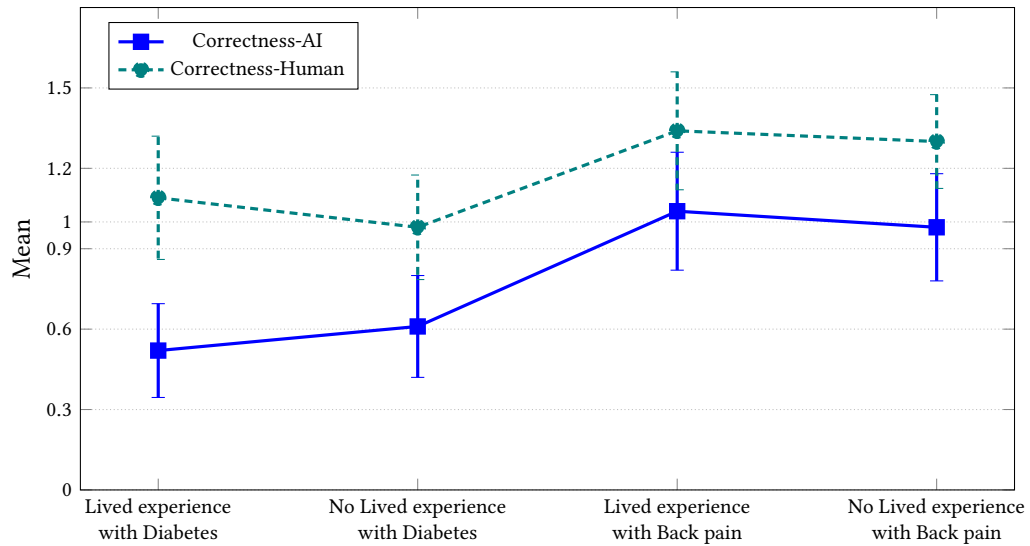

Fig. 6. Means of the Correctness-AI and Correctness-Human scores per Lived experience per Health condition. Error bars 95% CI.

To assess how people distinguish between AI-generated and human-written advice, we used ANCOVA. The independent variables in the model were Lived experience (Present vs Absent), Health condition (Diabetes vs Back pain), and Training (Present vs Absent). The outcome variable was Correctness-Overall. As trust might affect how people distinguish between human-written and AI-generated texts (Section 2.2) and the analyses pertinent to RQ1 (Section 4.1) identified a weak correlation between Trust and Correctness-Overall (Table 5), we used the Trust measure as a covariate. We checked the ANCOVA model for the necessary statistical assumptions.

The ANCOVA model did not identify any interaction effects between Training and Lived experience groups. Neither did it identify a significant association between the outcome variable and covariate. There was, however, a main effect of Health condition on correctly identifying whether the advice was written by human or generated by AI, $F(1, 243) = 32.05, p < .001, \eta_p^2 = .12$. The results indicate that people with and without the Lived experience of Diabetes scored significantly lower, $M = 1.42, SE = 0.08, 95\% \, CI[1.27, 1.57]$, in distinguishing between human-written and AI-generated health advice than people with and without Lived experience of Back pain, $M = 2.02, SE = 0.07, 95\% \, CI[1.87, 2.17]$ (with a mean difference of $dM = \pm 0.601$ significant at $p < .001, 95\% \, CI[+.392, +.810]/[-.810, -.392]$ and Trust evaluated at $M = 4.30$). This result undermines our hypothesis that people's experience of a chronic health condition might help them recognise the authorship of health advice online. At the same time, the finding indicates that the topic of health advice matters. It seems that potentially more complex and less common issues related to diabetes make it more difficult for people to distinguish the advice's authorship.

To gain insights about the outcome variable (Correctness-Overall) and learn whether the identified main effect exists in both contexts of distinguishing human-written or AI-generated texts, we also ran two ANCOVAs, looking at those scores independently, one with the outcome variable Correctness-Human and another with Correctness-AI. The effect was small yet significant for the human-written stimuli, $F(1, 213) = 6.85, p < .01, \eta_p^2 = .03$. Moreover, the effect was significant and larger when considering only AI-generated stimuli, $F(1, 210) = 19.75, p < .001, \eta_p^2 = .09$. Figure 6 visualises the means of the Correctness-AI and Correctness-Human variables over the combinations of levels of the Health conditions and Lived experience variables [4]. It illustrates how participants were able to perform more correct identifications of authorship for both AI-generated and human-written texts in the Back pain group compared with the Diabetes group, regardless of the presence or absence of Lived experience with the corresponding condition.

---

[3]Note that due to randomisation of stimuli in the study (Section 3.3.4), these two variables are not equally balanced for each participant, as each participant received a randomised number of human-written and AI-generated replies to evaluate (total three). However, randomisation was set equivalently for each between-subjects group to collect an equivalent structure of evaluations per group, and the analysis is group-based.

### 4.3 Qualitative indicators of advice authorship (RQ3)

In this section, we present findings from our qualitative analysis, focusing on our analysis of responses participants made to AI-generated health advice, to help us understand the reason for the observed difference in authorship evaluation effectiveness between the two Health conditions. Of the 24 codes developed across all responses, 23 were applied at least once to responses to AI-generated advice. Only one code (scaremongering) was applied to responses to human written text but not to AI-generated advice responses. Our analysis found people relying on quick reasoning [26] to help them decide on the authorship of the advice. In some instances, participants reported a feeling that helped them decide whether the advice was AI-generated or human-written, but they could not elaborate on why that feeling was occurring. For example, advice was described as needing to be "more human" or "robotic", without elaborating on what made the advice robotic-like or not human-like. However, other participants did elaborate by describing different characteristics of the advice that they used within their decision-making. These elaborative responses were core for our qualitative analysis, through which we developed 24 unique codes that were collapsed into three themes: (1) advice characteristics, (2) human characteristics, and (3) writing characteristics. When reporting our findings, we *italicise* where we mention a code.

*4.3.1 Advice characteristics.* The first theme is advice characteristics, including comments related to *perceived quality*, and the *level of detail* within the advice. When responding to AI-generated advice, 96 participants commented on an advice characteristic within their qualitative response. The most noticeable difference between the two Health conditions relates to the *level of detail*. A total of 34 participants mentioned *level of detail*, and out of this total, only 17.6% ($n = 6$) in the Diabetes group reported a lack of detail, whilst 50% ($n = 17$) reported a lack of detail in the Back pain group. For example, one participant, when asked what should be changed to make the text more human-like, wrote: "Provide more context to alternative options and suggest some options to consider." The next noticeable difference relates to the *perceived quality of the advice*, with 63 participants mentioning this in their responses. Of these, 19.4% ($n = 12$) in the Back pain group felt that the advice *includes bad or unclear advice* whilst fewer reported this in the Diabetes group (8.1%, $n = 5$). One example of this can be seen from one participant who wrote, "Seems to recite symptoms mainly rather than a balanced discussion of risks and benefits." Similarly, more people in the Back pain group (11.3%, $n = 7$) felt that answers did *not address the original question* compared to the Diabetes group (3.2%, $n = 2$). Finally, we found those in the Diabetes group (21.0%, $n = 13$) felt that the AI-generated advice was more *to the point and concise* compared to those in the Back pain group (4.8%, $n = 3$). In summary, participants in the Diabetes group reported more positive comments and fewer negative comments in their qualitative responses to AI-generated advice concerning advice quality and detail.

---

[4]The Training variable is omitted for brevity, given its null effect.

*4.3.2 Human characteristics.* The second theme is perceived human characteristics, including the level of *personalisation* and *emotional characteristics* within the advice. In total, 114 participants commented on human characteristics from AI-generated advice. Of these, the most prominent difference between the two conditions related to the *level of emotion*, with far more participants mentioning the *presence of emotional characteristics* within Diabetes advice (25.4%, $n = 29$) compared to Back pain advice (0.9%, $n = 1$). For example, one participant wrote "it just seems more informal and personal - there's a closeness and a certain intimacy to the language used". Supporting this finding, more participants felt that Back pain AI-generated advice *lacked emotion characteristics* (15.8%, $n = 18$), compared to Diabetes advice (6.1%, $n = 7$). When one participant was asked how the AI-generated advice could be made more human-like, they wrote: "Perhaps more understanding and empathy". There were also differences in how participants perceived the *level of personalisation* of the advice, with more participants considering Diabetes advice to be more personalised and tailored to the question being asked (11.4%, $n = 13$) compared to Back pain advice (1.8%, $n = 2$). When asked why they felt the AI-generated advice was human-written, one participant stated: "It seemed to be responsive to the patient's concerns, more so than [a] computer would be". We found more participants considering the Back pain advice (22.8%, $n = 26$) to be *too generic and textbook* compared to those in the Diabetes condition (15.8% $n = 18$) with one participant, when asked how the advice could be made more human-like, writing: "Lose the vagueness, waffle and arse-covering". In summary, participants in the Diabetes group reported more positive comments and fewer negative comments in their qualitative responses to AI-generated advice concerning the level of personalisation and the presence of emotional characteristics.

*4.3.3 Writing characteristics.* The final theme relates to writing characteristics within the advice. These characteristics relate to *writing style* (e.g., formal, informal), *grammar and language*, and *length*. In total, 76 participants commented on writing characteristics from AI-generated advice. For these, the most prominent difference between the two Health conditions relates to *writing style* with more participants considering the Back pain AI-generated advice (21.1%, $n = 16$) to contain *overly formal language* when compared to Diabetes advice generated by AI (10.5%, $n = 8$). For example, when one participant was asked how the AI-generated advice could be made more human-like, they wrote: "Make it less formal and more personal - feels like reading from an article rather than like a personal response". We found no difference in how participants across the two conditions spoke about the *use of grammar* and only a small difference in how people perceived the *length* of the advice they read. Finally, we found more participants in the Back pain (14.5%, $n = 11$) group felt that the AI-generated advice *contained repetition* compared to the Diabetes advice (6.6%, $n = 5$).

## 5 Discussion

Considering people's attitudes towards transparency and the relationship of such attitudes with trust in AI-generated advice (RQ1), our findings emphasise the importance of the relationship between trust and comfort in AI-generated health advice. Simultaneously, we show that transparency around factors that could influence advice given (financial and reputational information), as well as information on who is moderating advice, might be of importance. A positive correlation between trust and transparency indicates that trust in AI-generated health advice can be developed, but transparency about moderation provenance is needed.

The second research question focused on the investigation of whether people's experience with a given health condition, training on how to recognise AI-generated texts, and the health condition itself, could have an effect on their ability to distinguish advice authorship (RQ2). Although we did not find Lived experience and Training to have any effect on the ability to discern, we did find a difference between the two Health conditions. Our findings are broadly consistent with prior work within other contexts that shows people performing poorly at distinguishing authorship of AI-generated text [10, 38, 44, 46, 58].

Table 6 presents a summary of the results pertinent to the hypotheses addressing RQ1 and RQ2.

Table 6. Summary of research hypotheses (addressing RQ1 and RQ2).

| Label | Hypothesis | Result |
| --- | --- | --- |
| RQ1.H1 | Transparency attitudes towards health advice positively correlate with trust in AI-generated health advice. | Partially supported |
| RQ1.H2 | Transparency attitudes correlate with the ability to discern health advice authorship. | Partially supported |
| RQ1.H3 | Trust in AI-generated health advice correlates with the ability to discern health advice authorship. | Supported |
| RQ2.H1 | People who received Training on how to identify AI-generated health advice are detectably better at discerning the advice authorship, than those who did not receive such a Training. | Not supported |
| RQ2.H2 | People who have Lived experience with the health condition, to which the health advice is addressed, are detectably better at discerning the advice authorship, than those who do not have such a Lived experience. | Not supported |
| RQ2.H3 | There exists a detectable difference in the ability to discern health advice authorship between different Health conditions. | Supported |
| RQ2.H4 | People who both received Training and have Lived Experience are detectably better at discerning the advice authorship, than those who have either Training or Lived Experience. Those who have either Training or Lived Experience are also detectably better than those who have neither Training nor Lived Experience. | Not supported |

Lastly, we explored, through qualitative inquiry, characteristics within the generated advice that might help people discern the authorship of health advice (RQ3). In short, we identified that people tend to focus not only on content-related advice characteristics (e.g., level of detail) and writing characteristics (e.g., language) but also seek human characteristics (e.g., emotion, personalisation) in written health advice.

Below, we elaborate on our findings in more detail before discussing the implications of our results.

## 5.1 Transparency around AI-generated health advice

Our results add to the debate about AI transparency [32]. The answers to the transparency-related questions clearly indicate that, in the health advice context, people want to know more about AI authorship, who developed it, and who sponsored it. Such transparency is evidently needed since the ability to discern between human and AI-generated advice is poor.

The quantitative findings of the present research confirm the relationship between transparency and trust, particularly considering *Transparency of Private Moderation*, indicating that trust may increase with the increasing transparency around such moderation. In contrast, the relationship between trust and *Transparency of Societal Moderation* (i.e., when the moderator is assumed to have a public interest as a goal) being insignificant could be linked to the high trust in public services that have been recorded in the UK Office for National Statistics [59], particularly trust in NHS, reaching 80%. These differences also suggest that people may have different preferences for transparency of the reviewers moderating the health advice online, with AI-based moderators falling into the category of *Transparency of Private Moderation* where preferences may differ from *Transparency of Societal Moderation*.

**5.2 The unreliability of perceived honest signals**

The frequent use of human characteristics (e.g., the presence of emotion within the advice) as a means to discern human-written text suggests that people's decisions in this context might be based on automatic and fast reasoning — System/Type 1 thinking [19, 41]. Such reasoning would not be wrong if one were to follow the argumentation of Gigerenzer and Gaissmaier [25], stating that people tend to seek quick and easy ways of distinguishing between health advice, not because they are lazy but because, in the context of the study, this task might not have been important enough to apply more complex reasoning.

Within one of our Health conditions — Diabetes advice — the manipulative cues embedded within the AI-generated content resulted in flawed heuristic evaluations around the authorship of the advice, adding empirical weight to Hancock et al. [32]'s agenda on AI-mediated communication by showing the potential unreliability of linguistic cues that would ordinarily be considered reliable indicators of human-written text [14, 39, 54]. Whilst appearing honest, these cues were dishonest as they "indicate the existence of a quality that the signaler or environment does not actually have" [14] — in this case, human qualities. These dishonest signals were most likely used within participants' heuristic evaluations of the advice, resulting in a reduced ability to accurately discern authorship. Drawing on and supported by Walther's [79] hyperpersonal model of CMC, our findings show the lack of contextual cues amplifying these embedded cues within the advice. This finding supports and extends the findings of Jakesch et al. [39] into a different domain; their work found that people's inability to accurately discern AI-generated self-presentation authorship was partly due to flawed heuristics relating to grammar use and topic choice. In our study, we find these flawed heuristics developed similarly around grammar use (i.e., writing characteristics), but also show embedded cues linked to advice and human characteristics.

Donath [14] posits that signals lack reliability when they are sometimes honest and sometimes dishonest. Yet, to understand the reliability of a signal, receivers need to be able to discern where a signal has become dishonest. Within our study, participants were not informed of the honesty of the signals within the advice and so were unable to make this evaluation of reliability. With the human-like quality of their generated content, LLMs and other generative AI models are likely to continue producing content that is embedded with manipulative cues. Where this becomes more commonplace, there is a risk that this generated content negatively impacts the reliability of honest signals within human-written content. As services like ChatGPT do not intend their tools to be used to generate health advice, our work suggests the need for these tools to purposefully remove these unreliable signals leading to flawed heuristic evaluations. Moreover, where these tools are used to generate advice, we need to consider ways of enhancing transparency around their use to limit the risk of these tools reducing the reliability of signals within human-written advice across broader information systems.

**5.3 Implications**

*5.3.1 Restricting use entirely may not be appropriate.* Due to the potential risks of LLMs [22, 23, 65, 88], and the risk of their use within a health context [3, 17, 43, 74, 74], it could be argued that if an LLM can detect where health information is being sought, it should self-moderate to prevent health advice being provided to users to mitigate these risks. However, our work finds a large number of people reporting being happy to receive health advice generated by an AI chatbot (i.e., LLM-based), and prior work shows how, within the context of stigmatised conditions, advice from an AI chatbot might be preferred over human support [4]. Thus, we argue that moderation should have a non-restrictive format and instead build on transparency as a means of mitigating harm and enhancing user awareness. Our results indicate that such an approach might be more appropriate since almost all of our participants highlighted the importance of transparency around the provenance of LLM-based AI-generated advice. Therefore, we argue that LLM developers should consider how their models can be made more transparent to help users detect where LLMs have been used to generate health-related advice.

*5.3.2 Subtle and explicit embedded safeguards.* When generating a response using an LLM, there are often explicit cues within the generated output that can be used to help people recognise its source. For example, several published academic research papers include (presumably unintentionally) sections of text that are prefixed with these explicit cues such as the work of Zhang et al. [86] that includes the sentence "Certainly, here is a possible introduction for your topic" at the start of the introduction section of the paper.

Similarly, many LLMs embed explicit cues as safeguards into their products to help prevent their models from being used in ways they were not intended. For example, when prompting ChatGPT for health advice, the system embeds a text-based safeguard in the form of a prefix to the model output that reads: *"I am not a doctor, but I can try to provide some general information"* (Figure 1). Even where a model is generating text that requires fewer safeguards, there are often explicit signals within the prefix of a model's output, as in the above-mentioned example of the academic paper.

We argue that embedded safeguards like these are important to guide users, helping to prevent inappropriate use of LLMs (e.g., using them to diagnose health conditions or for legal advice), and to increase transparency around the provenance of such information (i.e., authorship). However, these safeguards are also easy to remove, i.e., they are low-cost signals that are easy to fake and so are not reliable [13, 14]. In our work, we show that without these safeguards in place (i.e., after they have been removed), people perform fast reasoning, relying on the potentially manipulative cues within the text to help them discern authorship. The findings from our research suggest that LLM developers should consider both explicit safeguards like the one shown in Figure 1, as well as more subtle safeguards that are harder to remove and, therefore, harder to fake and more reliable. For example, where it is detected that health advice is being sought, the generated output could be purposefully biased to remove dishonest signals such as human characteristics. While adversarial users could re-write the generated output to embed these characteristics back into the text, this is more costly in terms of time; Donath [14] argues that increases in cost (e.g., time/effort) will increase the reliability of signals.

*5.3.3 Provenance of advice.* Another way to assist people in making better decisions around trust and the use of AI-generated health advice can be by presenting them with sources of advice (e.g., references to where the information contributing to the advice came from). Overall, past research indicated that source might have an effect (e.g., credibility, trustworthiness, reputation) in the context of persuasion and communication of information [49]. In their meta-review, Ma and Atkin [50] showed the credibility of the source of information in the context of user-generated online health-related information matters. Of particular importance is a connection between where the information was posted and who created it. For example, the credibility of health information from blogs was perceived as high when the information was created by a layperson, but layperson-created information on generic websites was perceived as less credible. Similarly, past research indicates that online health information is perceived as more trustworthy when given by doctors, medical universities or governmental organisations [16]. Even when selecting the source of online health information, people prefer reputable organisations, considering the medical expertise [71] and branding (e.g., whether the company is recognisable, affiliated with a familiar institution, magazine, or similar) [87].

Although the above-mentioned research focused on human-generated health information, one could assume that similar source effects might be relevant in the context of health information generated by AI. To date, perhaps due to the novelty of LLMs, little attention has been dedicated to the provenance of AI-generated advice. Although some research around transparency of the source has been conducted (e.g., [49]), it was predominantly focused on informing users whether the information was generated by human or AI. The role of information sources within AI-generated advice only, has yet to be studied. There is, however, a growing interest in building technical solutions that can embed source information (i.e., references) into LLMs, to verify the received model's output and reduce model hallucinations [67], particularly in health-related contexts [84]. There are LLMs that attribute references to the generated text, but it remains unclear to what extent (e.g., wherein the text) the produced references support the AI-generated responses and their accuracy [84]. Our findings on transparency and cues that people use to discern AI-generated health advice, support the shift of the attention to research on sources of AI-generated health advice, both from a technical perspective, as well as from a CSCW and social computing perspective. These source-related findings also relate to the topic of authenticity, i.e., sources indicate who, beyond the LLMs, authored the advice. This is well aligned with growing industry interest in content provenance, for example, in initiatives such as The Coalition for Content Provenance and Authenticity (C2PA) [5].

### 5.4 Limitations and future work

We considered only two health conditions, and although this choice is somehow limited, it improved the study's feasibility, as the chosen health conditions are common among people of different demographics. Future research could seek to partially replicate the study across different health conditions.

The health-related questions and replies were generated by Reddit users. Inherent biases in these texts could have unknown effects on participants' performance or on the AI-generated texts. Similarly, all AI-generated texts were produced from a single model. Future studies could address this, involving a larger number of questions and replies from different sources and language models. Considering the advice required from Reddit, we could not assess whether the responses were 100% written by humans. However, half of the six responses used in the study were posted two to four years before the summer of 2023, when the availability of LLM-based AI-text-generating tools was more limited. Moreover, to avoid easily identifying the Reddit users, all questions and advice were altered by the study authors, making the texts more human-like.

We assessed participants' domain knowledge (health condition experience) through the proxy variable of whether they (or their loved ones) were affected by the condition. Although we believe this is likely a good proxy variable, it does not provide certainty on the domain knowledge level. For example, participants might not live with a health condition due to domain knowledge enabling them to take better precautions against it. Future work could assess the domain knowledge differently to validate our findings.

Diabetes texts used in our study were slightly longer than Back pain texts, which might have affected the evaluation and the final results of the study. However, the length of the texts corresponded to the original replies from Reddit, where there seems to be a discrepancy between the lengths of replies to questions related to the two Health conditions.

We used a single model (*GPT-3.5-turbo*) with a single set of prompt parameters. Other models, such as newer ChatGPT models (e.g., *GPT-4o*) as well as different prompt parameters, may result in different advice responses. Further work could consider the impact of different models, and different model parameters, on human-AI discernability.

Lastly, it is possible that qualitative results around writing characteristics might have been affected by Training on discerning human-written from AI-generated advice, in particular in responses related to grammar and language. However, not all participants have participated in the Training. Therefore, we believe such a potential priming effect would not be able to dominate the qualitative responses. Still, future research could investigate a comparison of how cues are perceived among participants who are and are not provided with the Training.

---

[5]https://c2pa.org/

## 6 Conclusions

While online community members are sometimes asked to help support the moderation of their own communities due to their knowledge of the community and its topic, we find that those with lived experience of a health condition are not effective at discerning between LLM-based AI-generated and human-written advice, even after training. As such, a community-based moderation approach to discern AI advice within online health communities may not be effective across all types of online health communities. Yet, differences in effectiveness at discerning authorship across health conditions due to more reliable (or the absence of less reliable) signals within health advice generated by LLM-based AI suggest that community-based moderation could be effective alongside improvements in the self-moderation of LLMs when they are instructed to generate health advice. Moreover, our findings indicate the need for a shift in research on the transparency of LLMs around health information. Considering the effect of the type of health condition identified in our study, we postulate that transparency, solely about whether the advice is generated by AI or not, is insufficient. Future research on LLMs should focus on embedding the usable provenance of health-related advice to enhance transparency and allow people to develop a better understanding of the complexity of health conditions.

## A Appendix

### A.1 Questionnaire measures

The questionnaire measuring the attitudes started after the main task. First, the instruction screen read,

"Below, we provide a number of statements regarding the role of transparency and trust when dealing with health advice online. Please indicate to what extent you agree or disagree with the statements."

*A.1.1 Attitudes towards transparency regarding AI-generated health advice.* The transparency measure followed the instructions and included 6 ad hoc statements, one of which was a multi-item statement including 6 items. Thus, the scale consisted of 11 items. All 6 statements and 6 items of the multi-item statement were presented in a randomised order.

(1) [*reversed item*] It is not important for me to know who is giving out health advice online.
(2) It would be helpful for me to know what other people think about health advice that I receive online.

(3) It is useful for me to know if a particular health advice I receive online is financially sponsored by a company or organisation, regardless of whether the advice comes from a human or an AI.
(4) I think it's helpful to know if a particular health advice I receive online is reviewed or moderated by:
    (a) Professional association (medical, academic, etc.)
    (b) National Health Service
    (c) Non-profit organisation (e.g., health charities)
    (d) For-profit organisation (e.g., private health providers)
    (e) Other members of an online health community
    (f) An AI health advice system
(5) It is important to me that whoever provides me health advice online also shares relevant personal or reputational details (e.g., affiliations, experience, education).
(6) If an AI was to provide me with health advice online, it is important to me that it declares information about who it was developed by?

The measurement scale included 7 points for each of the 11 items: Strongly disagree, Disagree, Somewhat disagree, Neither agree nor disagree, Somewhat agree, Agree, Strongly agree.

*A.1.2 Attitudes towards trust in AI-generated health advice.* The trust measure followed the transparency measure and included 6 single-item statements adapted from Nov et al. [58] and presented in a randomised order.

(1) I could trust AI-generated health advice about logistical questions such as health appointment times, and prescription requests.
(2) I could trust AI-generated health advice about preventative care, such as vaccines, or health screenings.
(3) I could trust AI-generated diagnostic advice about symptoms.
(4) I could trust AI-generated health treatment advice.
(5) AI-generated health advice can be a more trustworthy than a search engine to answer my health questions.
(6) AI-generated advice could help me make better decisions about my health.

The measurement scale included 7 points: Strongly disagree, Disagree, Somewhat disagree, Neither agree nor disagree, Somewhat agree, Agree, Strongly agree.

Additionally, the following ad hoc item was measured on the same scale in the trust attitudes section of the questionnaire stage:

(7) [*ad hoc item*] I would be comfortable receiving AI-generated health advice online.